\documentclass[aip,jcp,reprint]{revtex4-1} 
\usepackage{graphicx}

\usepackage{mathtools}
\usepackage{amsmath}
\usepackage{amsfonts}
\usepackage{graphicx}   
\usepackage{color}
\usepackage{ulem}
\usepackage{caption}
\usepackage{subcaption}
\usepackage{gensymb}

\usepackage{array}
\usepackage{booktabs}
\usepackage{multirow}

\usepackage{booktabs}

\begin{document}
\title{ Reweighted Autoencoded Variational Bayes for Enhanced Sampling (RAVE)}
\author{Jo\~ao Marcelo Lamim Ribeiro}
 \affiliation{Department of Chemistry and Biochemistry and Institute for Physical Science and Technology,
 University of Maryland, College Park 20742, USA.}
 \author{Pablo Bravo Collado}
 \affiliation{Departamento de Fisica, 
 Pontificia Universidad Catolica de Chile, Santiago 7820436, Chile}
 \affiliation{ University of Maryland, College Park 20742, USA.}
\author{Yihang Wang}
 \affiliation{Biophysics Program and Institute for Physical Science and Technology,
 University of Maryland, College Park 20742, USA.}
 \author{Pratyush Tiwary}
 \affiliation{Department of Chemistry and Biochemistry and Institute for Physical Science and Technology,
 University of Maryland, College Park 20742, USA.}

	\date{\today}
	
	\begin{abstract}
Here we propose the Reweighted Autoencoded Variational Bayes for Enhanced Sampling (RAVE) method, a new iterative scheme that uses the deep learning framework of variational autoencoders to enhance sampling in molecular simulations. RAVE involves iterations between molecular simulations and deep learning in order to produce an increasingly accurate probability distribution along a low-dimensional latent space that captures the key features of the molecular simulation trajectory. Using the Kullback-Leibler divergence between this latent space distribution and the distribution of various trial reaction coordinates sampled from the molecular simulation, RAVE determines an optimum, yet nonetheless physically interpretable, reaction coordinate and optimum probability distribution. Both then directly serve as the biasing protocol for a new biased simulation, which is once again fed into the deep learning module with appropriate weights accounting for the bias, the procedure continuing until estimates of desirable thermodynamic observables are converged. Unlike recent methods using deep learning for enhanced sampling purposes, RAVE stands out in that (a) it naturally produces a physically interpretable reaction coordinate, (b) is independent of existing enhanced sampling protocols to enhance the fluctuations along the latent space identified via deep learning, and (c) it provides the ability to easily filter out spurious solutions learned by the deep learning procedure. The usefulness and reliability of RAVE is demonstrated by applying it to model potentials of increasing complexity, including computation of the binding free energy profile for a hydrophobic ligand--substrate system in explicit water with dissociation time of more than three minutes, in computer time at least twenty times less than that needed for umbrella sampling or metadynamics.
\end{abstract}

	\maketitle
\section{Introduction}
\label{intro}
It is now routine to use molecular simulations in order to gain insight into difficult problems in the chemical, biological and material sciences. Such simulations have been facilitated via the development of more reliable molecular force-fields as well as powerful but accessible supercomputing resources. Despite these encouraging developments, however, it remains a challenge to simulate a large system over long timescales via brute-force computing. This is often the case because their energy landscapes contain a number of high barriers that separate various metastable states, trapping the simulation in limited parts of the landscape for extended periods of time. In order to solve this problem, several enhanced sampling methods have been proposed so as to accelerate the sampling of complex energy surfaces as well as facilitate the calculation of static and dynamic properties of rare events that are hard to sample.\cite{arpc_meta,tiwary2016review} In spite of how popular and useful these methods have become, however, the timescale problem has not yet been fully solved, and there remains a pressing need to develop newer and improved enhanced sampling methods.

Enhanced sampling methods themselves can be classified into different groups, as reviewed for instance in Ref. \onlinecite{tiwary2016review}. In one popular class of methods, the slow degree or degrees of freedom defining the reaction coordinate (RC) is/are first identified, so that fluctuations along the RC can then be enhanced leading to improved exploration of the energy landscape. Characteristic examples include metadynamics and umbrella sampling.\cite{arpc_meta} The common approach in such methods is to separate the aforementioned two steps so that first the RC is identified, either in an \textit{ad hoc} or a systematic manner; \cite{sgoop,metatica,ticameta} then, with the RC in hand, sampling is performed along the chosen RC. Some of the most recent work,\cite{sgoop,metatica,ticameta} however, have attempted to iterate between the steps, with sampling along a trial RC being used to ascertain an improved RC.

\begin{figure*}
  \centering
        \includegraphics[width=\textwidth,height=4in,keepaspectratio]{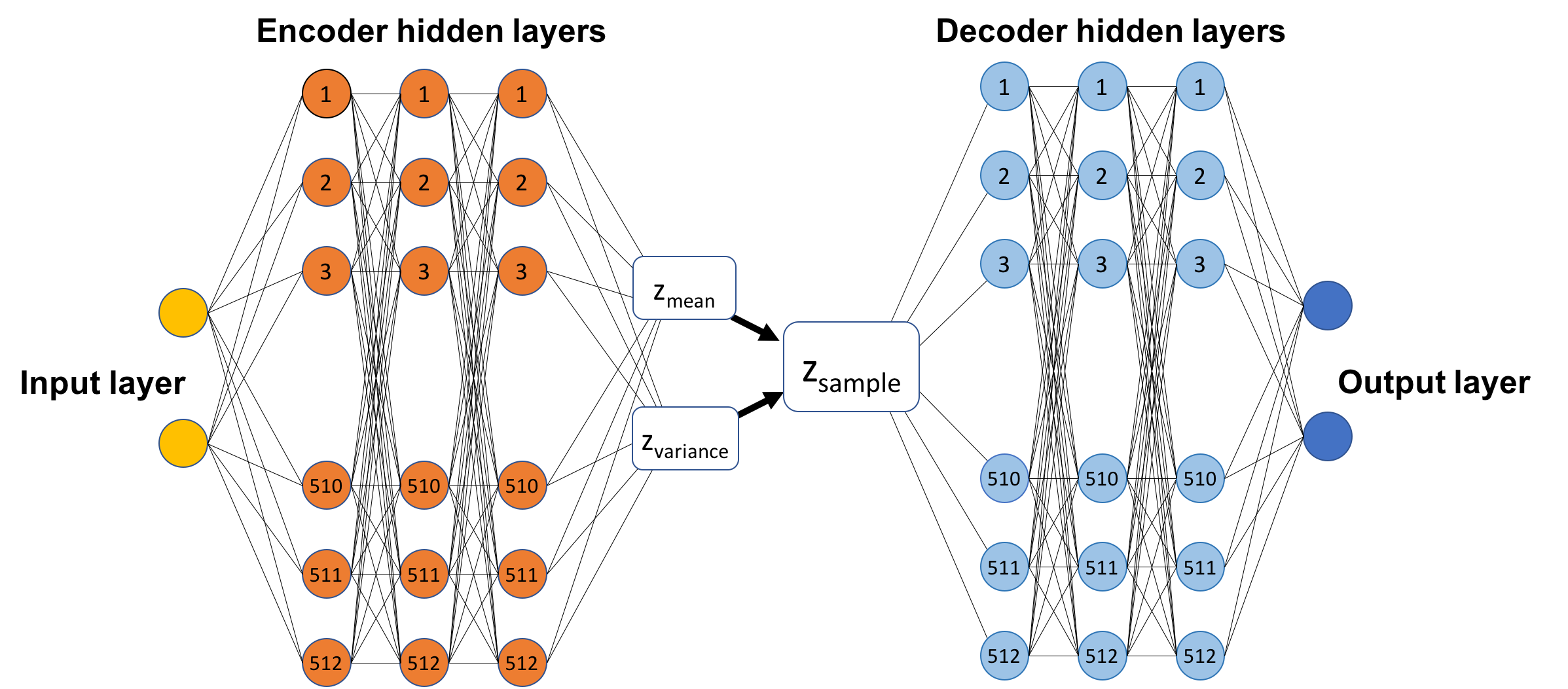}        
               \caption{A generic schematic illustration of the variational autoencoder model that also highlights the depth and width parameters of the deep neural networks specific to our work. The encoder neural network, in orange, maps a two-dimensional input into three sequential 512-dimensional vectors with the goal of learning \textit{two} one-dimensional latent variable parameters of a Gassian distribution, $z_{mean}$ and $z_{variance}$. The decoder neural network, in blue, maps a one-dimensional latent variable $z_{sample}$ taken from a Gaussian distribution into three sequential 512-dimensional vectors with the goal of reconstructing the original two-dimensional input. Please note that for the fullerene unbinding example both the input and output dimensions are three.}
\label{fig:architecture}
\end{figure*}

In this publication we will present a new enhanced sampling method that makes use of a state-of-the-art deep learning approach called variational autoencoder, and that combines, in a seamless manner, the identification of the RC together with the sampling of its distribution. The method iterates through rounds of molecular simulations, whose trajectories in terms of order parameters are fed to the deep learning module which then determines both the optimized latent variable representation for the RC as well as its probability distribution. Because such latent variable representations to the RC are devoid of physical interpretation, the method proceeds to locate an optimum but nonetheless physically interpretable RC from among a set of trial RCs via minimization of a suitably defined Kullback-Leibler divergence metric, also known as the relative entropy. Such an interpretable RC identified together with its distribution sampled from the molecular simulation then serves as the biasing protocol for the subsequent rounds of simulations, which are once again combined with the deep learning module but with the proper weighting accounting for the biased nature of the simulation -- hence the name Reweighted Autoencoded Variational Bayes for Enhanced Sampling (RAVE). The KL divergence or relative entropy has been previously used in the enhanced sampling community, albeit not in the context of leveraging deep learning. See for example Refs. \onlinecite{gobbo2017nucleation,gimondi2018co2,chaimovich2011coarse,shaffer2016enhanced}.

It has come to our attention during the preparation of this manuscript that several interesting enhanced sampling methods using deep learning techniques have become available in the recent literature.\cite{noe_vampnet,noe_timelaggedautoenc,ferguson_autoenc,sultan2018transferable} RAVE differs from these interesting methodologies in several respects. An important difference is that the recent methods continue to sample the RC distribution using an existing enhanced sampling approach while RAVE is independent of previous methods. Another crucial distinction is that while the methods in the literature continue to separate the biasing protocol into two steps, RAVE simultaneously identifies the RC as well as its unbiased probability distribution. Such simultaneous identification is not a question of simple aesthetics but it also allows RAVE to deal with the spurious local minima solutions to deep learning in a simple and coherent manner. This, in effect, provides a way to filter out the enhanced sampling results stemming from the misleading solutions. In this proof-of-concept paper, we summarize the main ideas behind RAVE and, in addition, demonstrate its usefulness on several model systems, including two analytical potentials as well as a a hydrophobic buckyball-substrate system in explicit water. All these systems have extremely high barriers (between 5 $k_BT$ and 30 $k_BT$) and using RAVE we demonstrate how we can obtain near-ergodic sampling and converged free energy profiles both accurately and efficiently. We conclude with a discussion of future directions as well as the challenges we see ahead. 

\section{Theory}
\label{theory}
\subsection{Variational Autoencoder}
\subsubsection{Overview}
RAVE makes use of the variational autoenconder (VAE) framework in order to model the MD trajectories. The theoretical foundation of the VAE is distinct from that of a traditional autoencoder,\cite{kingma2013auto,doersch2016tutorial,goodfellow2016deep} which is the most prevalent deep learning framework used thus far in enhanced sampling methods.\cite{noe_vampnet,noe_timelaggedautoenc,ferguson_autoenc} The VAE is a specific approach within the family of variational bayesian methods to modeling data generation, which is based upon the idea that the generative process consists of sampling from a prior distribution over a hidden latent space as well as from the likelihood:
\begin{equation}
p(\textbf{x}) = p(\textbf{x}|z) p(z)
\label{gen_model}
\end{equation}
In Eq. (\ref{gen_model}), $p(\textbf{x})$ is the generative model for the data \textbf{x}, while $p(z)$ is the prior over the hidden latent space and $p(\textbf{x}|z)$ is the likelihood. Notice that we have chosen to label the random variable representing the original high-dimensional datapoints as a vector, $\textbf{x}$, while the latent variable $z$ is left as a 1-dimensional random variable in order to reflect the restriction in this work that the latent variable representation to the RC be 1-dimensional. It is straightforward to generalize this restriction and it will be the subject of future work. 

Although as a generative model it suffices to have $p(z)$ and $p(\textbf{x}|z)$, as is clear from Eq. (\ref{gen_model}), the VAE does begin to resemble a traditional autoencoder since in order to train the VAE one first introduces a recognition model, $q(z|\textbf{x})$, in order to map the initial datapoints into the generative latent variable.\cite{kingma2013auto,doersch2016tutorial} The reason is that the actual VAE training objective (i.e. learning process), in practice, consists of maximizing a variational lower bound to the data's distribution, and not the distribution itself:\cite{kingma2013auto,doersch2016tutorial,goodfellow2016deep}
\begin{equation}
\mathcal{L} = \mathbb{E}_{z \sim q(z|\textbf{x})}\log p(\textbf{x}|z) - \mathcal{D}_{KL}(q(z|\textbf{x})||p(z)) \leq \log p(\textbf{x})
\label{gen_model_free_ene}
\end{equation}
In Eq. (\ref{gen_model_free_ene}), $\mathbb{E}_{z \sim q(z|\textbf{x})}$ denotes the expectation value of the likelihood when the latent variable is drawn from the recognition model while $\mathcal{D}_{KL}$ denotes the Kullback-Leibler divergence between the recognition model and the prior distribution. It is the training objective in Eq. (\ref{gen_model_free_ene}) that allows us to think of the VAE as being comprised of an encoder, $q(z|\textbf{x})$, mapping the original high-dimensional data into its low-dimensional latent space representation and a decoder, $p(\textbf{x}|z)$, mapping such a latent variable representation back into the original dataspace. The implementation of both the encoder and decoder within the VAE framework is done with the use of deep neural networks,\cite{kingma2013auto} which are a sequence of linear transformations that are passed through a non-linear function:\cite{goodfellow2016deep}
\begin{equation}
Z = \phi_{n} (A_n ... (\phi_{2}(A_{2}(\phi_{1}(A_{1} \bf{X} + b_{1})) + b_{2})) ... + b_{n})
\label{neural_net_transform}
\end{equation}
Eq. (\ref{neural_net_transform}) describes an encoder mapping an \textit{entire} dataset $\textbf{X}$ into a set of points in latent space $Z$ via several matrices of coefficients $A_{i}$, the vectors of coefficients $\textbf{b}_{i}$, and the non-linear functions $\phi_{i}$ through which the $i$th round of linear transformation is passed. Notice in addition that the depth of the neural network above is $n$, a user-defined feature representing the number of linear and non-linear combinations through which the data is passed. VAE decoders are implemented in an analogous fashion to Eq. (\ref{neural_net_transform}).

\begin{figure}
  \centering
        \includegraphics[height=6in]{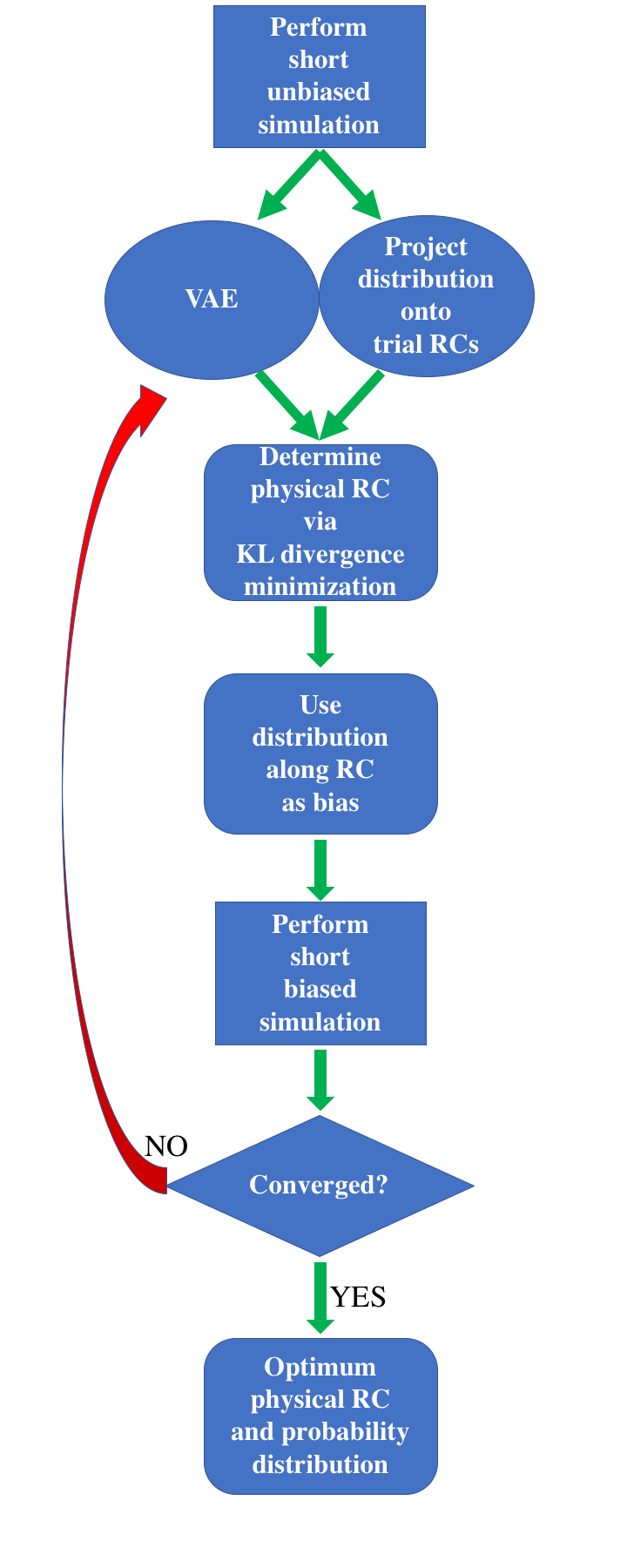}        
               \caption{A flowchart illustrating RAVE.}
\label{fig:flowchart}
\end{figure}

For the purpose of the work presented here we have chosen the VAE framework due to its aptness for learning reliable low-dimensional latent variable representations that can nonetheless capture the important features in the original data.\cite{goodfellow2016deep} In order to understand this, recall that in Eq. (\ref{gen_model_free_ene}) the variational lower bound $\mathcal{L}$ contains both an encoding and a decoding term: Maximization of $\mathcal{L}$ via an optimization algorithm will thus involve simultaneous learning of the encoder and decoder networks;\cite{goodfellow2016deep} the net result is that the VAE tends to arrive at a learned {low-dimensional} latent variable representation that can indeed capture the data's main features. In the context of this work, the latent variable representation will describe a low-dimensional manifold for the molecular simulation trajectories within configuration space. The approach that we take with the VAE, unlike other recent works using traditional or variational autoencoder methods for enhanced sampling,\cite{noe_vampnet,noe_timelaggedautoenc,ferguson_autoenc,sultan2018transferable} focuses on obtaining a high resolution mapping of the original molecular simulation data into its correct \textit{probability distribution} along the latent space. It is this focus on the probability distribution and not on the latent variable itself, that makes it unique among recent deep learning based enhanced sampling methods. Such an approach is inspired in part on some remarkable recent work on the Ising model, where the VAE framework was found to be capable of automatically learning both the block spin structure and also associated probability distributions, in the process recovering the findings commonly associated with the landmark Renormalization Group Theory.\cite{wetzel_ising}

\subsubsection{Neural Network Architecture}
It is important when using the VAE framework to make sure that the neural network architecture is suitable to the problem at hand. Interpreting neural networks as parametric function approximation machines,\cite{goodfellow2016deep} a suitable choice of the neural network architecture means defining an appropriate parameter space in which to learn a good function approximation. While approaches have been proposed to systematically optimize the network architecture, \cite{ferguson_autoenc} in general it remains the case that the choice of the neural network architecture is still the result of a great deal of trial and error. We provide in Fig. \ref{fig:architecture} a brief schematic illustration of some parameters for both the encoder and decoder used in the work here, while a more detailed breakdown of the neural network architecture is provided below.

\begin{enumerate}
\item{\textit{Input layer}}: The molecular dynamics (MD) trajectories, which for the two model potentials consists of 200,000 2-dimensional datapoints, while for the problem of fullerene unbinding consists of $\sim$6,000 3-dimensional datapoints.

\item{\textit{Encoder hidden layers}}: These first map each input MD datapoint into a sequence of three 512-dimensional vectors via the transformations $(\phi(A_{3}(\phi(A_{2}(\phi(A_{1} \bf{x} + b_{1}))+b_{2}))+ b_{3}))$, where $\phi$ is the ``exponential linear unit" (ELU).\cite{clevert2015fast} These then map the resulting 512-dimensional vector into two 1-dimensional parameters of a Gaussian distribution, the mean and variance, via the \textit{linear} transformation $A_{4} \bf{h}_{3} + b_{4}$.

\item{\textit{Decoder hidden layers}}: These first map a 1-dimensional latent variable, drawn from a Gaussian distribution using the parameters above, into a sequence of three 512-dimensional vectors via the analogous transformations $(\phi(A_{7}(\phi(A_{6}(\phi(A_{5} z + \bf{b}_{5}))+b_{6}))+ b_{7}))$, with $\phi$ the ELU function. Then maps the resulting 512-dimensional vector into the space of the original MD dataset via the transformation $\phi(A_{8} \bf{h}_{7} + b_{8})$, where $\phi$ is either the sigmoid or tanh functions.
\end{enumerate}

The implementation and training of the neural network just described was done using a high level deep learning library named Keras.\cite{chollet2015keras} The optimization algorithm that we have used during training was the RMSprop, a variation of the stochastic gradient descent, with a learning rate of 0.005. All other parameters were left at their default values as implemented in Keras. Training was performed for 100 epochs except in the later rounds of the fullerene unbinding work due to the rather large weights from the biased simulations forcing the training to be over a longer period of time.

\subsection{Reweighted Autoencoded Variational Bayes for Enhanced Sampling (RAVE)}
%\subsubsection{Overview}
We now proceed to describe RAVE, which is also summarized through a flowchart in Fig. \ref{fig:flowchart}. Although the description will now be specific to MD simulations, all of what follows will hold as well for Monte Carlo. In order to initiate RAVE, a short MD simulation is run, but for a realistic system with barriers $\gg k_B T$ it is quite probable that the simulation will remain trapped in its initial state. Feeding the data from this unbiased MD simulation into the VAE, the deep neural network learns a concise 1-dimensional latent space \textit{z} within which the higher dimensional MD trajectory is embedded, as well as the probability distribution along this space. However, while the latent space definition from the VAE is a continuous and differentiable function of the original input variables, it lacks a clear physical interpretation. Here, then, the emphasis is shifted from the latent space definition itself to its probability distribution. RAVE, screening for various linear and in principle non-linear combinations of input order parameters thus identifies the RC $\chi$ as the one whose probability distribution as sampled in the input MD trajectory most closely resembles the one learned from VAE. The Kullback-Liebler divergence metric is used as a measure of this resemblance between the two probabilities, which is defined as follows:
\begin{equation}
\mathcal{D}_{KL}(P(z)||P(\chi)) = \sum_{i} P^{u}(z_{i}) \text{ log} \dfrac{P^{u}(z_{i})}{P^{u}(\chi_{i})}
\label{kl_div}-
\end{equation}
In Eq. (\ref{kl_div}), $P^{u}(z)$ is the unbiased distribution that the encoder within the VAE framework learns, $P^{u}(\chi)$ is the unbiased distribution stemming from the projection of the MD data onto the combinations of input order parameters, and the summation \textit{i} is over the 1-dimensional gridded spaces $z$ and $\chi$ that have been both normalized and discretized to the same number of bins. The candidate distribution $P^{u}(\chi)$ that minimizes Eq. (\ref{kl_div}) thus identifies the RC given the current amount of sampling. Next, RAVE takes the RC as well as its probability distribution to construct the bias, $V_{bias}(\chi)$, for a next round of MD simulation, which is defined as follows:
\begin{equation}
V_{bias}(\chi) =   k_{B}T \text{ log}P^{u}(\chi) = k_B T \text{ log} \langle \delta (\chi -\chi (t))\rangle
\label{first_bias}
\end{equation}
where the ensemble average is performed over the unbiased trajectory $\chi(t)$. The bias in Eq. (\ref{first_bias}) is in the spirit of conformation flooding or metadynamics, \cite{grubmuller,mccarty2015variationally,arpc_meta,meta_laio} with the additional advantage that the task of identifying the RC and a suitable bias is now  combined and automated. With this bias potential RAVE runs a biased MD simulation using the total potential, $V_{MD} = V_0({\bf{R}}) + V_{bias}(\chi(\bf{R}))$, where $V_0(\bf{R})$ is the unbiased potential energy of the system given as a function of the configurational coordinates $\bf{R}$. See Sec. \ref{gencomm} for further details.

It is important to remember that the MD simulation whose data is fed into the next RAVE iteration is now biased. Thus although in principle RAVE proceeds to use Eq. (\ref{kl_div}) to screen among a number of trial RCs, it must first produce the unbiased probability distribution from a biased simulation. This is done through proper reweighting of the simulation so that both the VAE as well as the projections of the MD data onto the trial RCs incorporate the correct statistics. Before projection each MD datapoint now carries a weight given by:
\begin{equation}
w = e^{V_{bias} / k_{b}T}
\label{md_weight}
\end{equation}
The unbiased probability is obtained from a biased simulation through the use of the simple reweighting formula for importance sampling:
\begin{equation}
P^{u}(\chi) = \frac{\langle w\delta (\chi -\chi (t))\rangle_b }{\langle w\rangle_b} 
\label{unbiased_biased_projection}
\end{equation}
where the subscript $b$ denotes sampling under a biased ensemble with weights from Eq. (\ref{md_weight}). With Eq. (\ref{unbiased_biased_projection}), we have obtained from a biased simulation the denominator in Eq. (\ref{kl_div}).
The VAE as well needs to account for the weighted data and RAVE implements the reweighting of the VAE within the \textit{reconstruction} loss function, such that the actual learning of the latent dimension will incorporate the correct statistics:
\begin{equation}
\sum_{i} w_i^2(\textbf{x}_{i} - \textbf{y}_{i})^2 = \sum_{i} (w_i\textbf{x}_{i} - w_i\textbf{y}_{i})^2
\label{vae_weight}
\end{equation}
In Eq. (\ref{vae_weight}), \textbf{x} denotes an individual datapoint belonging to the original MD simulation data, \textbf{y} denotes the reconstruction of the individual datapoint from the VAE, $w_{i}$ are the weights given in Eq. (\ref{md_weight}), and the summation $i$ extends over the total number of points in the entire MD dataset $\left[\textbf{x}_{1}, \textbf{x}_{2}, \textbf{x}_{3}, ... \right]$. Eq. (\ref{vae_weight}) is just the mean squared error between the MD data and its VAE reconstruction with the proper weights attached to each of the datapoints. It amounts to performing the stochastic gradient descent (i.e. the learning process) in a configuration space with the reweighted statistics. Implementation of the weighted reconstruction loss function leads to the unbiased probability distribution, which is the numerator in Eq. (\ref{kl_div}). With both unbiased probabilities RAVE then proceeds to use Eq. (\ref{kl_div}) and locate the optimum biasing parameters, $\chi$ and $P^{u}(\chi)$, for another round of biased MD. From here, RAVE can now enter into another iteration and it continues in a loop until desired thermodynamic observables, in the case of this work the free energy, is converged.

\begin{figure*}
  \centering
        \includegraphics[height=2.6in]{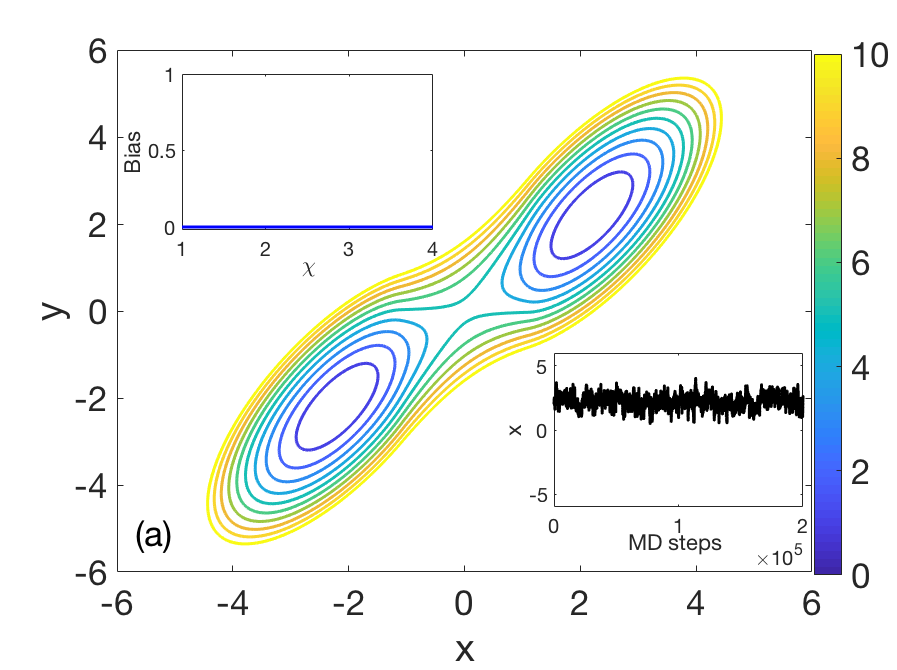} 
        \includegraphics[height=2.6in]{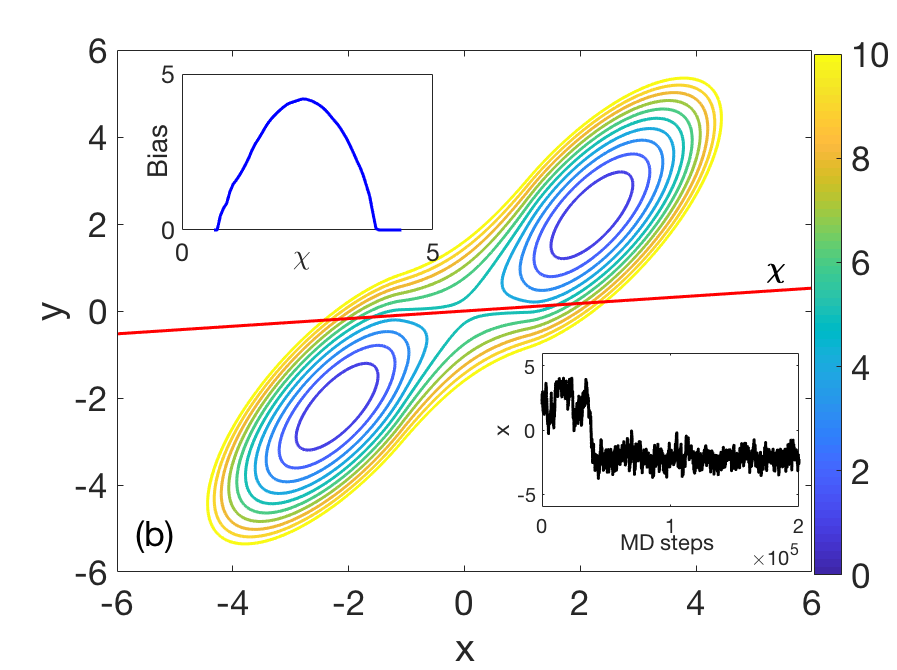}
        \includegraphics[height=2.6in]{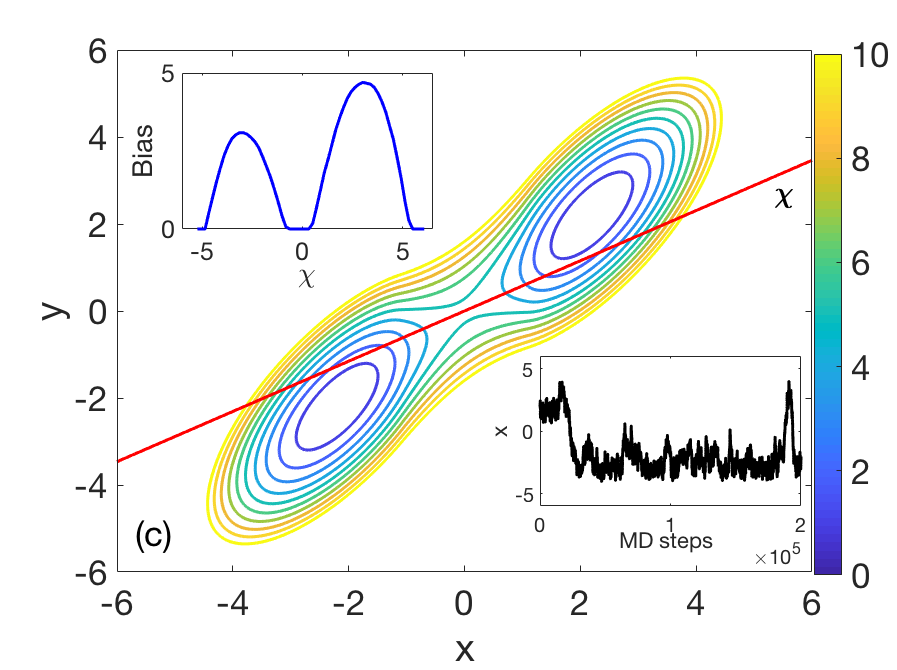}
               \includegraphics[height=2.6in]{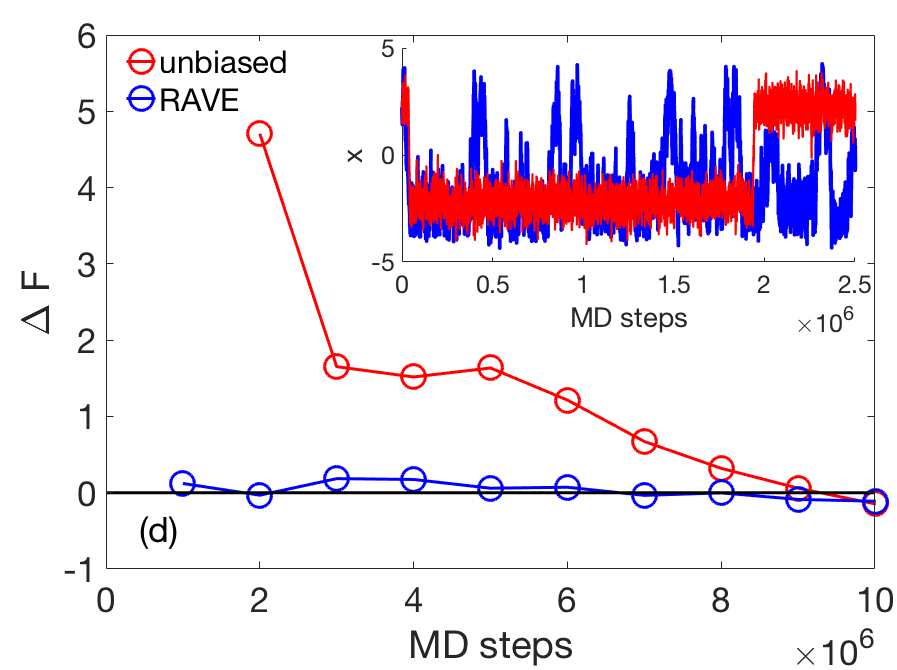}
\caption{(a-c) Countour plots of the Szabo-Berezhkovskii model two-state potential, $V(x,y)$, with the respective top and bottom inserts representing the bias potential and $x-$coordinate as a function of MD timesteps associated with (a) the unbiased simulation and (b-c) the 1st and 2nd biased simulations. Red lines represent the current RC $\chi$. All energies are in units of $k_BT$, with each contour line denoting a 1 $k_BT$ interval. (d) The differences in the free energy between the two available wells, with the top insert representing the $x-$coordinate as a function of MD timesteps comparing the final biased MD simulation with the unbiased simulation.}
\label{fig:szabo}
\end{figure*}

\begin{figure*}
  \centering
        \includegraphics[height=2.6in]{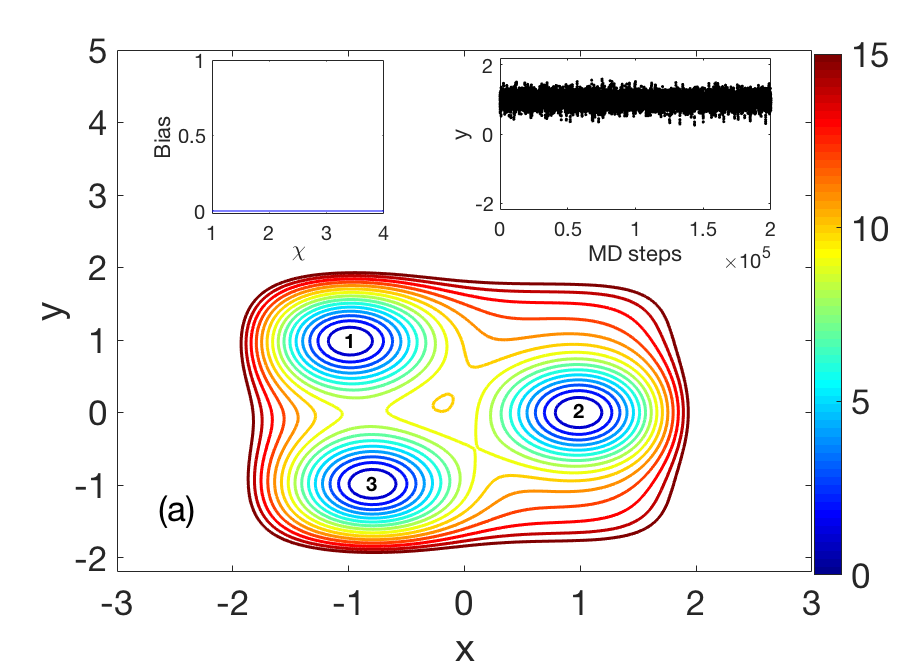} 
        \includegraphics[height=2.6in]{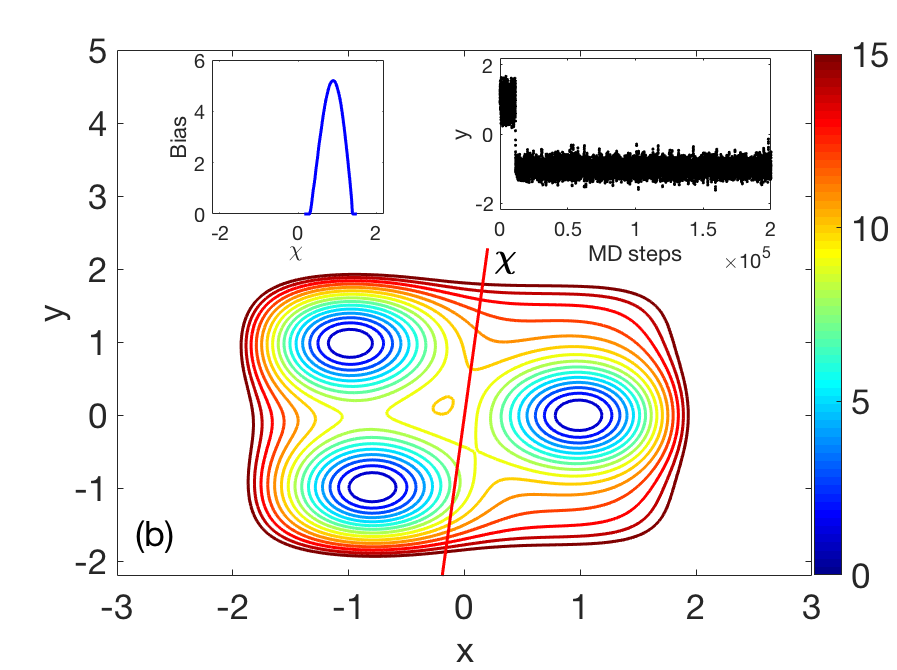}
         \includegraphics[height=2.6in]{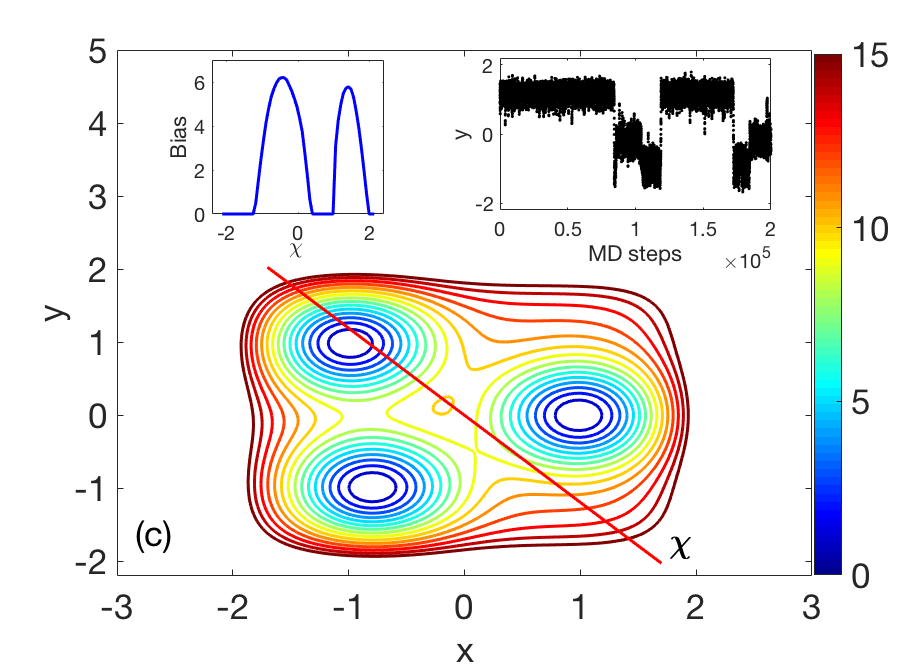}
               \includegraphics[height=2.6in]{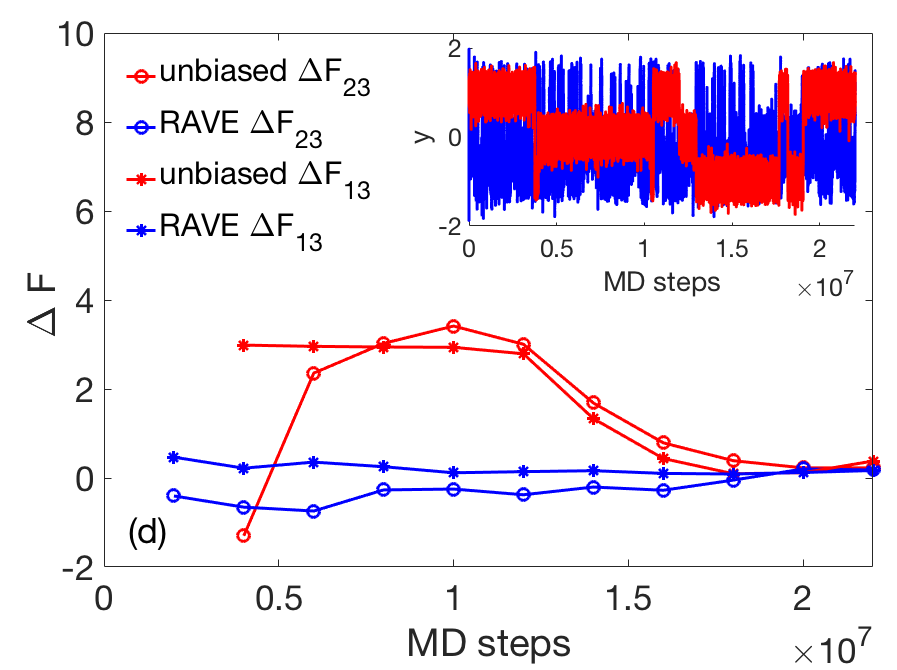}
\caption{Countour plots of the three-state model potential, $V(x,y)$. The top inserts on the left and on the right represent the bias potential and $y-$coordinate as a function of MD timesteps associated with (a) the unbiased simulation and (b-c) the 1st and 5th biased simulations. Red lines represent the current RC $\chi$. All energies are in units of $k_BT$, with each contour line denoting a 1 $k_BT$ interval. (d) The differences in the free energy between pairs of available wells, with the top insert representing the $y-$coordinate as a function of MD timesteps comparing the final biased MD simulation with the unbiased simulation.}
\label{fig:multistate}
\end{figure*}

% The VAE is designed in a way to free the training algorithm to learn the latent variables whose semantics are not known \textit{a priori}. Deep neural network for not  just where data lives, but how data moves. Markovianity is implemented in the loss function. 
% \begin{enumerate}
% \item where we different from other recent autoencoder/neural net methods: we prioritize the probability distribution of the latent variable rather than the definition of the latent variable
% \item task of learning RC definition and RC probability distribution has been traditionally treated separately. Here we combine them in a seamless manner
% \item VAE learns RC and distribution of RC at same time. VAE is unsupervised and learns number of states as well on the fly
% \item VAE already gives us perfect bias but it is as a function of a probabilistic RC. VAE gives the ideal low-dimensional code and its probability distribution. We will find a cheap analytical approximation to this code.
% \item In SGOOP (see Section. \ref{intro}) we found best code by iterating over combinations of OP with maximal spectral gap - here we do it by iterating over combinations of  OP whose probability density is closest to the Q(z) learnt by VAE, then use this density as next round of bias
% \item can this bias be tuned to allows use of acceleration factors. See Fig. \ref{fig:algo} for a description of the algorithm.
% \end{enumerate}

\section{Results}
\subsection{Model Two-State Potential}
For our first illustrative example, we have applied RAVE to the well-studied Szabo-Berezhkovskii potential,\cite{szabo_anisotropic} whose contour plot is given in Fig. \ref{fig:szabo}. Here, as well as for the three-state model potential, Newton's second law of motion was integrated for a canonical ensemble using Langevin dynamics\cite{bussi_langevin} with an integration timestep of 0.01 units at temperature $k_{B}T=1$. In this low temperature regime, a short unbiased MD simulation does not escape from the well where it was launched from and instead the simulation just oscillates about the initial minima. Using this unbiased MD data as input for the VAE, the line with $\theta=5\degree$ as shown in Fig. \ref{fig:szabo}(b) was determined to be the optimum RC for this 0th, unbiased RAVE iteration.

%{Notice from the bottom insert in Fig. \ref{fig:szabo}a that the entire MD simulation is constrained to the PES well whose x-coordinate is positive.} 

%Keep in mind that since the distribution along these lines are the result of projections of an MD dataset that is $\sim$featureless due to it being trapped in a single PES well, the KL divergence comparison shows a large amount of invariance with respect to $\theta$ and is often $\sim$flat. Two spikes appear in the plot of the KL divergence versus $\theta$, however, these peaks centered about $\theta=135\degree$ and $315\degree$, which are known a priori as being the worst available RCs for the Szabo-Berezhkovskii potential. The presence of such KL divergence spikes indicate that even prior to escape from the initial PES well the method can learn to discard the worst available RCs. In order to understand this, notice that the MD data has oval shaped PES wells that project a distribution onto the lines about $\theta=135\degree$ and $315\degree$ that turn out to be of different widths than the learned distribution about the latent dimension. Relevant plots of the KL divergence versus $\theta$ will be provided in supporting materials.

Now that both a RC as well as a probability distribution have been identified, we can generate a bias using Eq. \ref{first_bias} in order to run another short, but biased, MD simulation. As can be seen in the bottom insert of Fig. \ref{fig:szabo}(b), the biased MD simulation samples regions of the PES that went unexplored during the unbiased simulation. Once the MD simulation transitions out of the initial PES well it then becomes trapped in the second well. The reason is that the 0th RAVE iteration leads to a single-peak bias that acts on a single PES well, in essence lowering the barrier height in just the \textit{forward} direction. Now that we have the 1st biased MD simulation we proceed in a manner analogous to before: The optimum RC after this 1st RAVE iteration was then determined to be along $\theta=30\degree$ while the bias that was generated from the optimum distribution was two-peaked. Note that this RC is in excellent agreement to the analytical result of Ref. \onlinecite{szabo_anisotropic} as well as the calculation made through other methods.\cite{anisod_sgoop} Looking at Fig. \ref{fig:szabo}(d), we can see that using these updated biasing parameters in another short MD simulation leads to effective ergodicity in the dynamics as seen through several quick transitions as well as extremely fast convergence of the free energy difference between the two basins relative to an unbiased MD run of same duration. Two rounds of RAVE, then, was all it took to achieve ergodicity.
\subsection{Model Three-State Potential}
Next, we applied RAVE to enhance the sampling of a three-state model potential, whose contour plot is given in Fig. \ref{fig:multistate}, and which is defined as: 
\begin{equation}
\begin{split}
V(x,y) = &-12 \left \{ e^{-2\left(x+1 \right)^2 -2\left(y-1 \right)^2   }   \right \} \\
& -12 \left \{ e^{-2\left(x+0.8 \right)^2 -2\left(y+1 \right)^2   }   \right \} \\
& -12 \left \{ e^{-2\left(x-1 \right)^2 -2y^2   }   \right \} \\
\end{split}
\label{model_three_state}
\end{equation}
A short unbiased MD simulation at temperature $k_BT=1$, with other simulation parameters similar to those described for the previous potential, is again unable to escape the initial well but it can be used in conjunction with the VAE in order to get the distribution along the latent dimension. With this learned latent variable distribution the optimum RC was determined to be along $\theta=85\degree$, while the bias that was generated was single-peaked (Fig. (\ref{fig:multistate}b)). Using these as the biasing parameters in a new MD simulation led to a quick transition from the initial well, showing that with just one RAVE iteration the simulation has overcome the 8 $k_{B}T$ barrier that separates two of the three wells. 
%{The reason for the biased MD simulation connecting the initial PES well with the well located in the third quadrant is that the line with $\theta=85\degree$ has the correct slope for connecting these specific PES wells.}

The optimized RC and the bias so-constructed for different rounds of RAVE are given in Fig. \ref{fig:multistate}. As can be seen from this figure, after five RAVE iterations, the trajectory becomes significantly more ergodic, and the free energy difference between the basins also converges extremely quickly as compared to the unbiased MD.

\begin{figure*}
  \centering
        \includegraphics[height=2.6in]{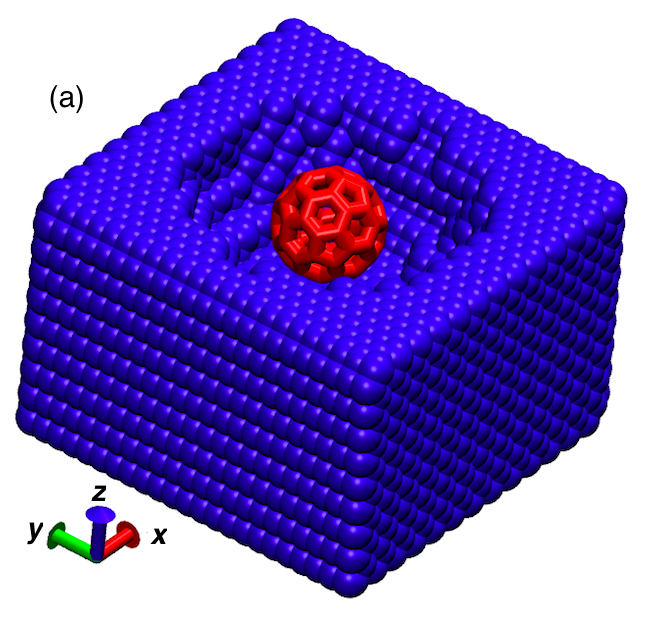} 
        \includegraphics[height=2.6in]{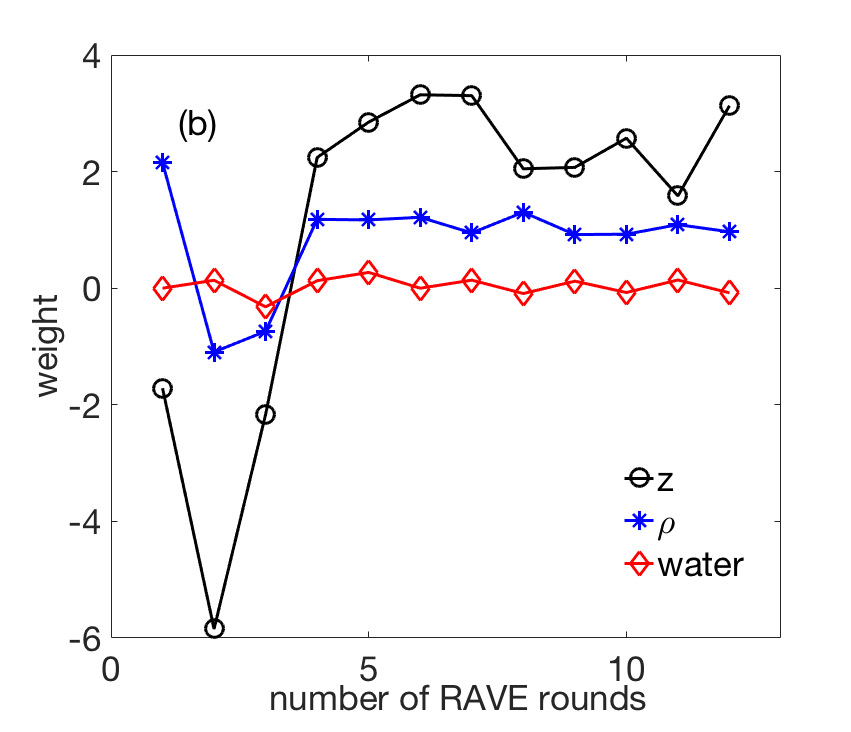}
               \includegraphics[height=3.5in]{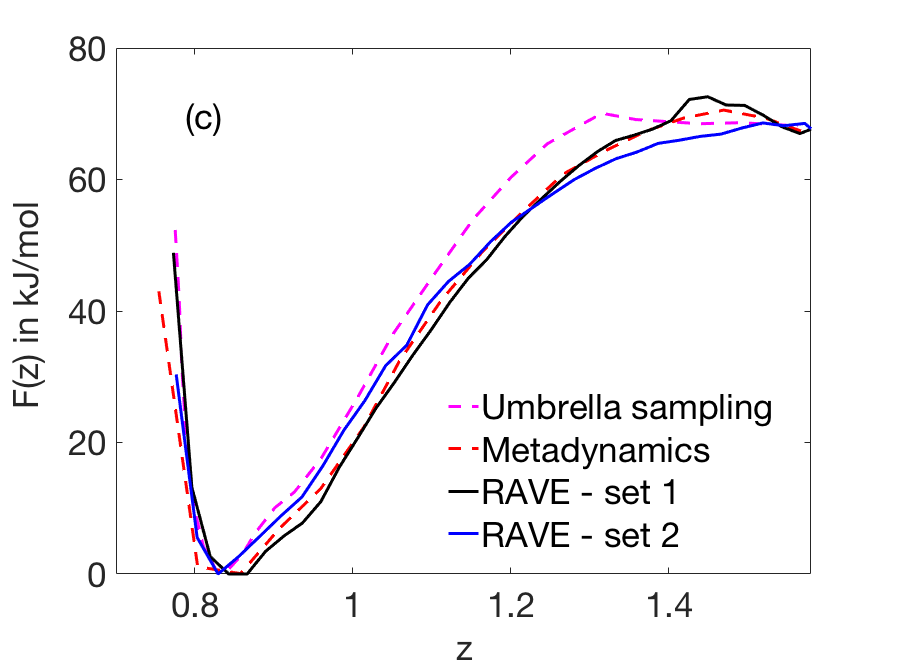}
\caption{(a) Hydrophobic ligand-cavity system in explicit water. Cavity and ligand atoms are colored blue and red respectively. The water molecules are not shown for clarity. Axes have been marked. See Ref. \onlinecite{sgoop_fullerene,mondal_fuller} for further details of the system set-up.  (b) Weights in reaction coordinate versus number of RAVE rounds carried. Convergence was obtained after around 10 rounds. (c) Free energy profile along $z$ as obtained using RAVE (black and blue solid lines using two different final rounds), two-dimensional umbrella sampling (magenta dashed line) and one-dimensional metadynamics (red dashed line). The last two profiles have been taken from Ref. \onlinecite{sgoop} and we refer to that publication for details of the simulations. }
\label{fig:fullerene}
\end{figure*}

%With the first biased MD dataset in hand, the subsequent RC "picked" from among the set of available candidates is the line with $\theta=90\degree$, which similar to the previous RC also connects the wells in the second and third quadrants. In addition, a two-peak bias is generated. Using these current biasing parameters in another biased MD simulation, several quick transitions between the two PES wells connected by the RC are observed until the simulation escapes into the third PES well. Once the simulation escapes into the third PES well, it remains trapped there due to the absence of a bias in that region of the PES. At this point, after two RAVE iterations, all PES stable states have been visited; we continue to iterate, however, using $F_{max}$ as guide, until the biased MD simulation becomes $\sim$ergodic, with constant transitions between all available states, as well as a converged free energy profile. Such a set of convergence criteria is met after a 5th iteration in which RAVE picks out a RC for $\theta=130\degree$. Fig. \ref{fig:multistate}c summarizes the results of this 5th (and final) biased MD simulation while the final convergence parameters are provided in Fig. \ref{fig:multistate}d.

\subsection{Hydrophobic Ligand-Cavity System in Explicit Water}
We now tackle the unbinding of a fullerene-shaped ligand from a host cavity in explicit water at a temperature of 300 K, illustrated in Fig. \ref{fig:fullerene}(a). This, as well as related systems, have been widely studied over the years \cite{mondal_fuller,sgoop_fullerene} in order to understand a range of physical processes such as nanoassembly and drug unbinding. Here, the system that is used is identical to the one in Ref. \onlinecite{sgoop_fullerene} and we refer to that publication for details. The fullerene is free to move in any direction. Unlike in the case of the two previous model potentials, where the test for RAVE was to achieve ergodic sampling of multiple wells, the test now is whether RAVE is robust enough to surmount a very high barrier of $\sim$30 $k_{B}T$ corresponding to a residence time of 200 seconds,\cite{fullerene} which would correspond to an unbiased MD simulation of more than 1,000,000 years even with the best available supercomputing resources. A second question we ask is whether RAVE can reproduce the free energy profile for this system, and if so, how does the computational time compete with methods such as umbrella sampling and metadynamics.

The MD trajectories here are of 0.5 ns duration in each round and comprise a time-series of the three variables: (i) $z$, or the z-component of the fullerene-cavity separation, (ii) $\rho = \sqrt{(x^2+ y^2)}$, or the axial fullerene-cavity separation and (iii) $w$, the solvation state of the cavity. These three variables are defined in detail in Ref. \onlinecite{sgoop_fullerene}. The optimized RC is kept of the form $c_z z + c_{\rho} \rho + c_{w} w$ where the three coefficients are the weights of the respective order parameters in the RC.

For this system, after about 10 RAVE iterations, it was found that the RC, as measured by the weights described above, converges to a value very similar to the one reported by Tiwary and Berne in Ref. \onlinecite{sgoop_fullerene}. Namely, the weight of the solvation state variable almost disappears entirely, while the highest weight corresponds to the $z$ variable followed by the $\rho$ variable (Fig. \ref{fig:fullerene}(b)). After 22 rounds of RAVE, we obtained a bias strong enough to cause unbinding of the ligand in multiple independent short MD runs. The free energy profiles as function of $z$ so-obtained from two independent final RAVE rounds, started with randomized positions and velocities, are provided in Fig. \ref{fig:fullerene}(c). There is clear agreement with umbrella sampling and metadynamics in terms of the binding free energy and the entire binding free energy profile. Furthermore, we note that the net computer time used for RAVE was at least 20 times less than that reported for umbrella sampling and metadynamics in Ref. \onlinecite{fullerene}. 

This example demonstrates clearly that apart from obtaining an accurate free energy profile in much less computer time than at least two other enhanced sampling methods, we are also able to extract a physically relevant RC from the deep learning procedure. Namely, our RC captures the role of steric and solvation effects that has been highlighted in previous works.\cite{sgoop,sgoop_fullerene} It will be very interesting to apply this procedure to more realistic ligand unbinding systems and see what information we can extract there.

\subsection{General Comments on the Usage of RAVE}
\label{gencomm}
Here we would like to state some heuristics and observations that were found efficient and useful while implementing RAVE. Deep learning protocols are often prone to getting trapped in a local minima, thus giving misleading solutions.\cite{goodfellow2016deep} These spurious solutions can often all correspond to similar loss functions, which can be quite deceptive. In fact, enhanced sampling algorithms such as tempering have been used to accelerate the convergence of deep learning modules to the true solution! \cite{bengio2013deep} In such a case, one faces a chicken versus egg problem, and it is not trivial to know which is the correct solution. To deal with such a significant issue, ours is a two-fold approach. First, under the constraint that the bias must be zero in the regions that were not sampled, since we simply have no information about these regions, we rank the various solutions as per the maximal bias recorded. In the case of the bias given in Fig. \ref{fig:multistate}(b), this would be $\sim$5 $k_BT$. A large fraction of spurious solutions from VAE were found to have much lower values for this maximal bias metric, and on this basis ruled out. Some additional observations regarding this heuristic is covered in the Discussion section. Second, in the uncommon case that multiple solutions are found to pass the first test with a similar metric, all are then used in the next round of biased MD simulation and the one with maximally enhanced exploration of the free energy landscape is selected for the next round of RAVE. It is current work in our group to make these physically motivated criteria further robust and rigorous.

\section{Discussion}
\label{discussion}
In this work we have proposed RAVE, an iterative scheme that uses the VAE deep learning framework to enhance sampling in MD simulations. RAVE is based on the idea that the probability distribution of the latent space can be taken as the most relevant feature learned from the VAE as opposed to the precise definition of the latent variable itself. The motivation in shunning the precise latent variable that the deep learning framework learns is two-fold: First, it is not intuitive, and mapping it to close approximations can be desirable when it leads to an increase in physical intuition; 
second, for a rough potential energy surface the true RC is a complicated non-linear function of the numerous configurational coordinates which we do not seek to replicate. We wish instead to find a relevant \textit{feature} of the true RC that can provide a good measure for how well the cheaper and intuitive RC proxies approximate the true RC. In other works such as SGOOP by Tiwary and Berne, \cite{sgoop,sgoop_fullerene,tiwary_biotin} the approximation is quantified by how large is the spectral gap of the projected dynamics, and in other methods \cite{metatica,ticameta} some other dynamical property is taken as this metric. Here we choose as our benchmark the probability distribution that the VAE deep learning framework learns. It is possible that these approaches could also be combined through the use of a more refined objective function, which is something we are in the midst of exploring. %Since the VAE learned latent variable can in principle describe complex non-linear functions, if we map an MD dataset into its learned distribution along the latent dimension, we arrive at a feature derived \textit{in} the complicated and non-linear dimension that can describe the true RC. Thus as the distribution along some RC proxies approach that which is along the latent dimension, those simpler curves will have best captured the most important but complicated aspects of the true RC itself as well as the true RC's distribution.
It is important to keep in mind that RAVE also allows for the possibility of matching the VAE probability distribution more accurately by the use of more complex RCs. The end result is that RAVE allows one to choose just how much intuition and computational cost to sacrifice when defining the RC while when building a method on top of the latent variable itself one is forced to deal with the aforementioned complicated non-linear variables lacking in a great deal of intuition.

Our heuristic of choosing the VAE solution with maximal bias is inspired by the maximixing spectral gap approach of the SGOOP method from Ref. \onlinecite{sgoop}. Roughly speaking, under a constant diffusivity approximation, a representation with deeper energy basin, or higher maximal bias, will have highest first passage time out of the basin, and thus highest spectral gap.  For example, the various local minima solutions obtained from a round of VAE could be screened using the Maximum Caliber\cite{caliber1,dixit2015inferring} based framework of SGOOP to decide which one is more likely to be the global minima. We are exploring this intriguing connection between RAVE and SGOOP and hope to report our findings in future work.

In all the applications considered here, RAVE was found to be much faster than unbiased MD, several orders of magnitude so, and for the hydrophobic ligand-cavity unbinding system it was $\sim$20 times faster than even metadynamics or umbrella sampling. Of course, the use of VAE adds computational overhead but with the use of GPUs and especially for larger high-barrier systems this overhead should be minimal. In summary, in this work we have introduced a new deep learning based enhanced sampling method that gives both the reaction coordinate and its probability distribution at the same time, without having to recourse to additional enhanced sampling methods. By iterating between rounds of deep learning and molecular dynamics, we are able to obtain converged estimates of thermodynamic observables with minimal prior intuition and limited computational workload. {The systems considered here have 2 or 3 order parameters only -- our initial tests as well as work by others\cite{ferguson_autoenc} suggest that this number could be easily increased significantly. Another area we are pursuing actively involves implementing temporal identity in the protocol, in the spirt of time-lagged autoencoders for example.} We are hopeful this method will add a new tool in the exploration of complex molecular systems plagued with rare events.

\vspace{.2in}

\textbf{ACKNOWLEDGMENTS}\\
PT thanks Dr. Steve Demers for suggesting the use of variational autoencoders. The authors thank Sebastian Wetzel for sharing variational autoencoder code with them. The authors acknowledge the University of Maryland supercomputing resources (http://hpcc.umd.edu) made available for conducting the research reported in this paper. 

\bibliography{tiwary_references}

	\end{document}